\documentclass[11pt,twoside]{article}

\usepackage{asp2006}
\usepackage{epsf}
\usepackage{lscape}

\begin{document}
\title{How `heredity' and `environment' shape galaxy properties}
\author{Gabriella De Lucia}
\affil{Max Planck Institut f\"ur Astrophysik, Postfach 1317, D-85748 \\
Garching bei M\"unchen, Germany}

\begin{abstract}
  In this review, I give a brief summary of galaxy evolution processes in
  hierarchical cosmologies and of their relative importance at different
  masses, times, and environments.  I remind the reader of the processes that
  are commonly included in modern semi-analytic models of galaxy formation, and
  I comment on recent results and open issues.
\end{abstract}

\section{Introduction}   
It has been known for a long time that the local and large--scale environment
play an important role in determining many galaxy properties\footnote{First
  indications of a correlation between the galaxy {\it type} and the {\it
    environment} can be found in the {\it The Realm of Nebulae} by E. Hubble
  (1936).}.  Disentangling the processes responsible for the observed
correlations has proved difficult and it remains unclear whether the observed
relations are imprinted during formation or by physical processes at work in
dense environments.  In recent years, the subject has received much impetus
from the completion of large spectroscopic and photometric surveys at different
redshifts \citep[e.g.][]{kauffmann04,cucciati06,cooper06}.  Historically,
however, theoretical and observational studies trying to assess the role of
environment on galaxy evolution have been focused mainly on galaxy clusters.
The primary reason for this is the practical advantage of having many galaxies
in a relatively small region of the sky and all approximately at the same
redshift.  This allows efficient observations to be carried out, even with
modest fields of view and modest amounts of telescope time.  It should be
noted, however, that in order to establish that physical processes related to
the cluster environment are indeed playing a role, it is necessary to compare
the evolution of similar galaxies in different environments (i.e. in the
clusters and in the `field'). In addition, it is worth reminding that galaxy
clusters represent a {\it biased} environment for evolutionary studies.  In the
current standard cosmogony, clusters originate from the gravitational collapse
of the highest peaks of primordial density perturbations, and evolutionary
processes in these regions occur at an accelerated pace with respect to regions
of the Universe with `average density'.  Clusters do have another related
important drawback: they are {\it rare} (they only contain about 10 per cent of
the cosmic galaxy population at the present day, and even a lower fraction at
higher redshift).  Finally, there is one element that is often overlooked in
classical discussions about environment: according to the current paradigm for
structure formation, dark matter collapses into haloes in a bottom-up fashion.
Small systems form first and subsequently merge to form progressively larger
systems.  As structure grows, galaxies join more and more massive systems,
therefore experiencing different `environments' during their life-times.  In
this context, it is clear that both `heredity' (i.e. the initial conditions)
and `environment' (i.e. subsequent physical processes that galaxies experience
during their life-times) do play a role in shaping the observed galaxy
properties and in determining the observed environmental trends.

\section{Physical processes}

A comprehensive review of early and recent theoretical work on galaxy evolution
and environment would easily fill an entire volume.  In the following, I will
therefore limit myself to an overview of the various physical processes, and of
their relative importance at different masses, times, and environments.  I will
remind the reader of the processes that are commonly included in modern
semi-analytic models of galaxy formation, and comment on recent results.

{\bf Mergers :} Galaxy mergers and more generally strong galaxy-galaxy
interactions, are commonly viewed as a rarity in massive clusters because of
the large velocity dispersion of the system.  They are certainly more efficient
in the infalling group environment and may therefore represent an important
`preprocessing' step in the evolution of cluster galaxies.  In this
perspective, mergers are important over the entire life of a galaxy cluster: at
early times when the cluster is first collapsing, and still at later times in
the outskirts of the cluster, as it accretes groups from the field.  Numerical
simulations \citep[see][and references therein]{Mihos04} have shown that close
interactions can lead to a strong internal dynamical response driving the
formation of spiral arms and, in some cases, of strong bar modes.  The
axisymmetry of these structures leads to the compression of the gas that can
fuel starburst/AGN activity.  Simulations have also shown that sufficiently
close encounters can completely destroy the disc, leaving a kinematically hot
remnant with photometric and structural properties that resemble those of
elliptical galaxies.

\begin{figure}[!ht]
 \plottwo{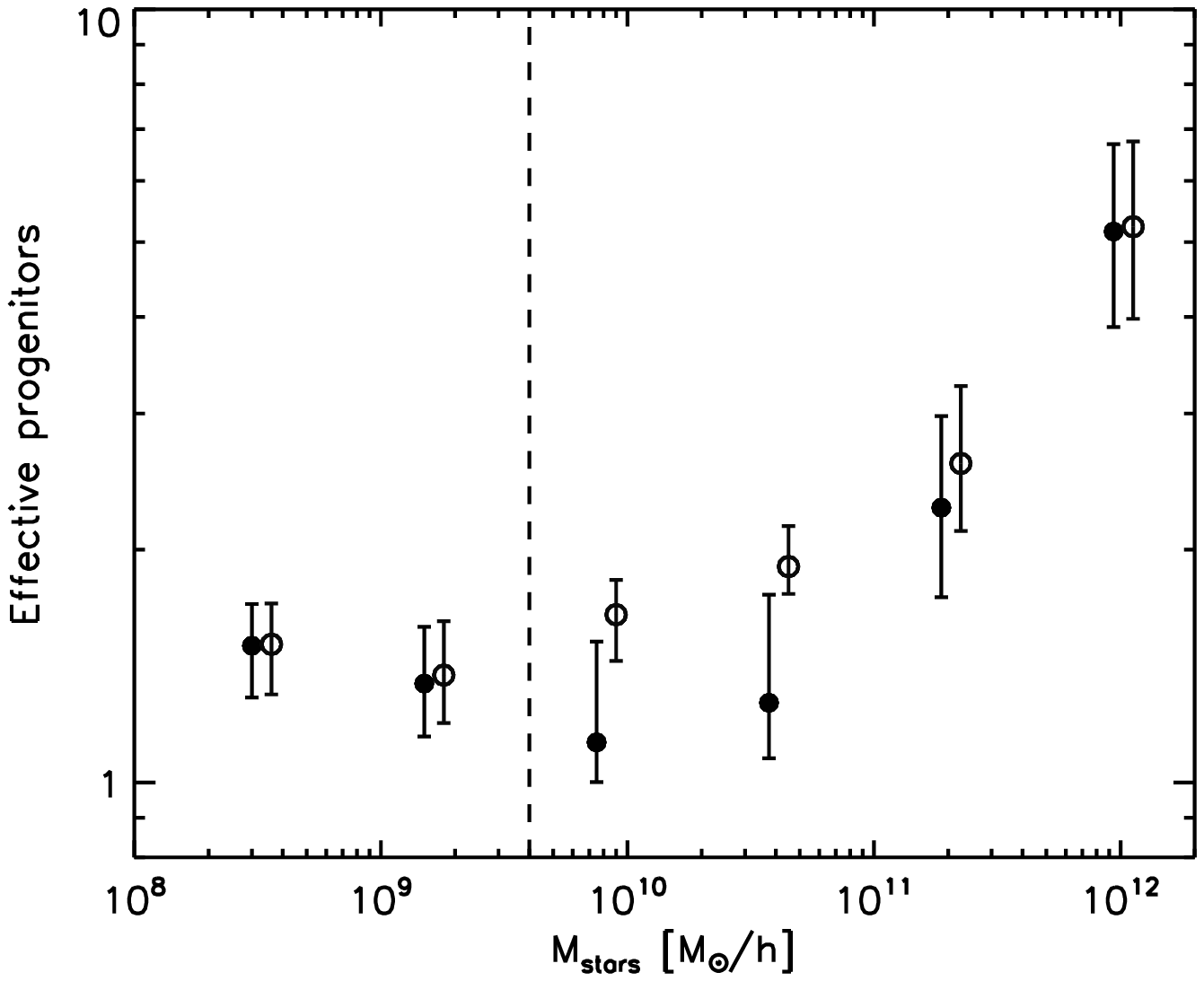}{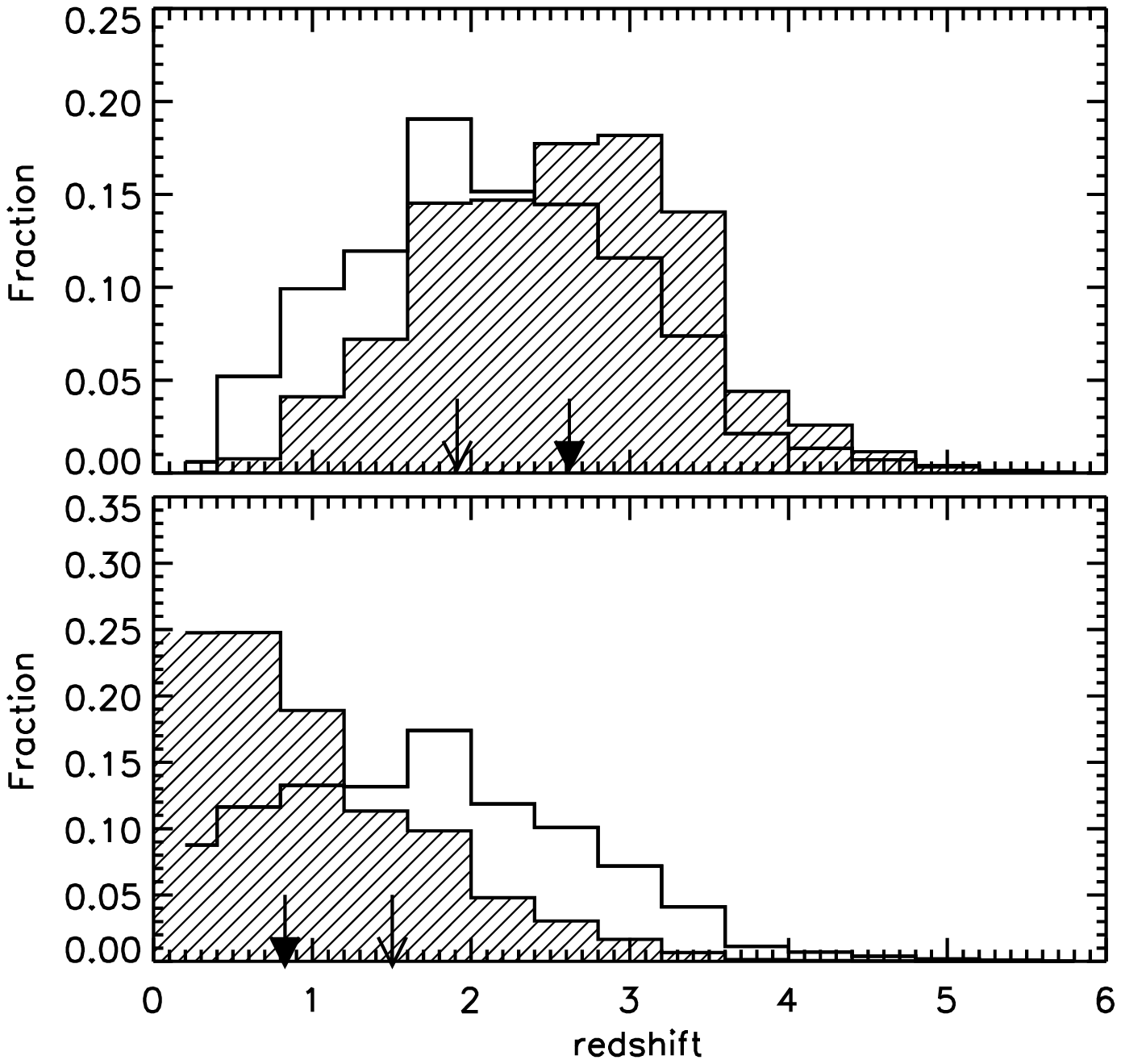}
 \caption{{\itshape Left:\/} Effective number of progenitors as a function of
   galaxy stellar mass for model elliptical galaxies. Symbols show the median
   of the distribution, while error bars indicate the upper and lower
   quartiles. Filled and empty symbols refer to a model with and without a disc
   instability channel for the formation of the bulge. {\itshape Right:\/}
   Distribution of formation (top panel) and assembly redshifts (bottom panel).
   The shaded histogram is for elliptical galaxies with stellar mass larger
   than $10^{11}\,{\rm M}_{\sun}$, while the open histogram is for all the
   galaxies with mass larger than $4\times10^{9}\,{\rm M}_{\sun}$. Arrows
   indicate the medians of the distributions, with the thick arrows referring
   to the shaded histograms.\citep[From][]{delucia06}}
 \label{fig:fig1}
 \end{figure}
Mergers are intrinsically included in standard semi-analytic models of galaxy
formation and represent the main channel for the formation of bulges.  In the
hierarchical galaxy formation scenario, more massive galaxies form through the
mergers of smaller units and larger systems are expected to be made up by a
larger number of progenitor galaxies.  It is therefore interesting to ask how
the number of progenitors varies as a function of galaxy mass.  In our recent
work \citep{delucia06}, we have investigated this issue by defining for each
galaxy an {\it effective number of stellar progenitors}.  This essentially
represents a mass-weighted counting of the stellar systems that make up the
final galaxy (see original paper for more details) and therefore provides a
good proxy for the number of significant mergers required to assemble a galaxy.

The left panel of Fig.~\ref{fig:fig1} shows how the effective number of
progenitors varies as a function of galaxy mass for model ellipticals.  Filled
circles show results from our `standard' model which includes a disc
instability channel for the formation of bulges.  Empty symbols, indicate
results from a model in which this channel is switched off.  The vertical
dashed line indicates the threshold above which our morphology classification
can be considered `robust' (see original paper for more details).  As expected,
more massive galaxies are made up of more pieces.  The number of effective
progenitors is, however, less than $2$ up to stellar masses of $\simeq
10^{11}\,{\rm M}_{\sun}$, indicating that the formation of these systems
typically involves only a small number of major mergers.  Only more massive
galaxies are built through a larger number of mergers, reaching up to $\simeq
5$ for the most massive systems.  The right panel of Fig.~\ref{fig:fig1} shows
the distribution of `formation' (top panel) and `assembly' redshifts (bottom
panel) of model ellipticals.  The formation redshift is defined here as the
redshift when $50$ per cent of the stars that end up in ellipticals today are
already formed, while the assembly redshift is defined as the redshift when
$50$ per cent of the stars that end up in ellipticals today are already
assembled in a single object.  The right panel of Fig.~\ref{fig:fig1} shows
that more massive galaxies are {\it older}, albeit with a large scatter, but
assemble {\it later} than their lower mass counter-parts. The assembly history
of ellipticals hence parallels the hierarchical growth of dark matter haloes,
in contrast to the formation history of the stars themselves.  Data shown in
the right panel of Fig.~\ref{fig:fig1} imply that a significant fraction of
present elliptical galaxies has assembled relatively recently through purely
stellar mergers, in agreement with recent observational results
\citep[e.g.][]{dokkum05}.

{\bf Harassment :} Galaxy harassment is a process that is not usually included
in semi-analytic models of galaxy formation.  The process has been discussed in
early work on dynamical evolution of cluster galaxies
\citep[e.g.][]{richstone76}, and has been explored in some detail using
numerical simulations by \citet[][]{FS81}.  These early studies showed that
repeated fast encounters coupled with the effects of the global tidal field of
the cluster, can drive a strong response in cluster galaxies - results that
were confirmed by later and better simulations \citep{moore98}.  Numerical
simulations indicate that the efficiency of this process is largely limited to
low-luminosity hosts, due to their slowly rising rotation curves and their
low-density cores.  For this reason, it is believed that harassment might have
an important role in the formation of dwarf ellipticals or in the destruction
of low-surface brightness galaxies in clusters, but it is less able to explain
the evolution of luminous cluster galaxies.

{\bf Gas stripping :} Galaxies travelling through a dense intra-cluster medium
suffer a strong ram-pressure stripping that sweeps cold gas out of the stellar
disc \citep{gunn72}. Although the gas is tenuous, a large pressure front builds
up in front of the galaxy because of its rapid motion.  Depending on the
binding energy of the gas in the galaxy, the intra-cluster medium will either
be forced to flow around the galaxy or will blow through it removing some of
the diffuse interstellar medium.  Related mechanisms are thermal evaporation
\citep{cowie77} and viscous stripping of galaxy discs \citep{nulsen82}, that
occur when ram-pressure is not effective. In this case, turbulence in the gas
flowing around the galaxy entrains the interstellar medium resulting in its
depletion.  Unlike the physical mechanism discussed before, gas stripping does
not affect galaxy morphology.  At least not directly.  But once star formation
is halted in a disc, this can fade significantly, the bulge-to-disc relative
importance can change, and therefore the galaxy morphology can appear
different.  The effect of ram-pressure stripping has been discussed only in a
couple of studies using semi-analytic techniques \citep{okamoto03,lanzoni05}.
These conclude that the inclusion of this additional physical process causes
only mild variations in galaxy colours and star formation rates.  This happens
because the stripping of the hot gas from galactic haloes (see below)
suppresses the star formation so efficiently that the effect of ram-pressure is
only marginal.

\begin{figure}[!ht]
 \plotone{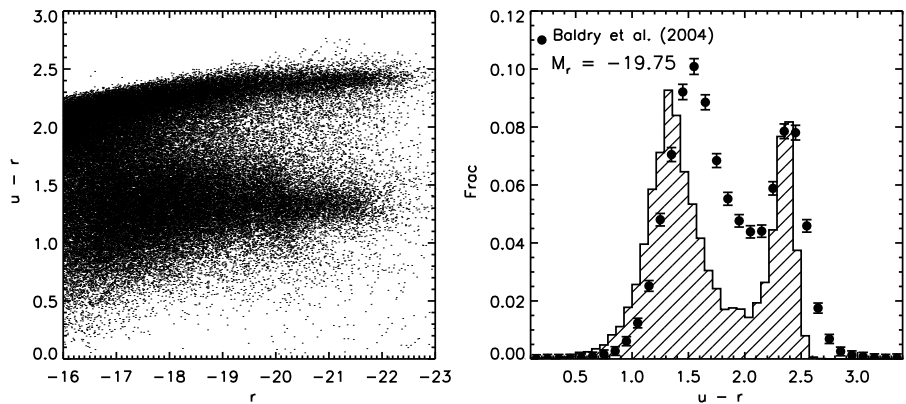}
 \caption{{\itshape Left:\/} u-r vs r distribution for galaxies in a model with
   relatively efficient supernovae feedback. {\itshape Right:\/} Colour
 distribution (shaded histogram) of model galaxies in the magnitude bin
 $[-19.7, -19.8]$.  Data points show observational measurements from
 \citet{baldry04}.} 
\label{fig:colour}
\end{figure}

{\bf Strangulation :} Current theories of galaxy formation suggest that when a
galaxy is accreted onto a larger structure, the gas supply can no longer be
replenished by cooling that is suppressed \citep{larson80}. This process has
been given the quite violent name of `strangulation' (or `starvation' or
`suffocation') and it represents one important element of semi-analytic models
of galaxy formation.  It is common to read in discussions related to these
physical mechanisms that strangulation is expected to affect a galaxy star
formation history on a quite long timescale and therefore to cause a slow
declining activity.  This is not what happens in practice.  If this process is
combined with a relatively efficient supernovae feedback (and this is the case
in many recent models), galaxies that fall into a larger system consume their
cold gas very rapidly, moving onto the red-sequence on quite short time-scales.
The left panel of Fig.~\ref{fig:colour} shows the u-r vs r colour-magnitude
relation for such a model. The right panel shows the colour distribution of
model galaxies in the magnitude bin $[-19.7, -19.8]$.  The figure clearly shows
that the `transition' region does not appear to be as well populated as
observed.  It also appears that there is a clear excess of faint red satellites
with respect to observational data (see also Balogh, this volume).

{\bf AGN heating :} Since the milestone paper by \citet{white91}, it has been
realized that some physical process is needed to suppress cooling flows that
otherwise would produce too many massive and luminous galaxies at odds with
observational results.  Early semi-analytic models introduced ad-hoc
prescriptions to suppress cooling flows in haloes above a critical
mass.  Modern models have included more accurate and physically motivated
prescriptions and have confirmed that AGN heating is indeed important to
reproduce the exponential cut-off at the bright end of the galaxy luminosity
function \citep[e.g.][and many others]{croton06,bower06}.  These prescriptions
are necessarily very schematic and not well grounded in observation.  Still
much work needs to be done in order to understand exactly how and when AGN
feedback is important.

\begin{figure}[!ht]
 \plotone{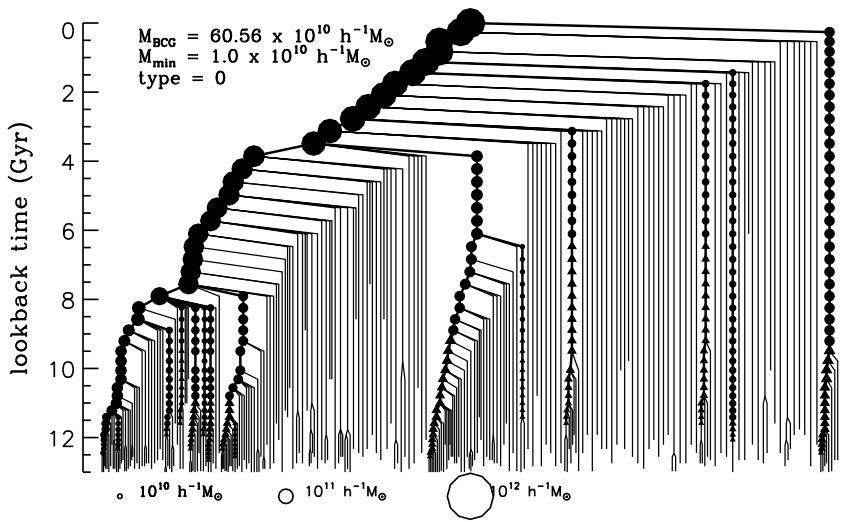}
 \caption{Merger tree of a central galaxy of a cluster-sized halo. The area of
   the symbols scales with the stellar mass.  Only progenitors more
  massive than $10^{10}\,{\rm M}_{\odot}\,h^{-1}$ are shown with symbols.
  Circles are used for galaxies that reside in the FOF group inhabited by the
  main branch. Triangles show galaxies that have not yet joined this FOF
  group. \citep[From][]{deluciablaizot06}.} 
 \label{fig:tree}
 \end{figure}

{\bf Cannibalism :} Early theoretical studies have discussed the role of
cannibalism due to the dynamical friction in the formation of brightest cluster
galaxies \citep{OstrikerTremaine75,White76}.  This early work was however not
successful due to the use of a simplified cluster model (this relates to my
discussion in Sec.~1).  In the now standard paradigm of structure formation,
clusters assembled quite late, through the merging of smaller systems. In this
perspective, cooling flows are the main fuel for galaxy formation at high
redshift, in dense and lower mass haloes.  This source is removed at lower
redshift, possibly due to AGN feedback.  Galaxy-galaxy mergers, as discussed
above, are most efficient within small haloes with low velocity dispersion.
They are indeed driven by dynamical friction, but it is the accretion rate of
the galaxies into the proto-cluster, along with the cluster growth itself that
regulates and sets the conditions for galaxy merging.  This is illustrated very
nicely in Fig.~\ref{fig:tree} which shows the merger tree of a central galaxy
of a cluster-sized halo \citep[for details and for a colour version of the
figure, see][]{deluciablaizot06}.  Fig.~\ref{fig:tree} shows another important
point: in the context of the hierarchical paradigm for structure formation, the
full history of a galaxy is described by its complete merger tree.  Whereas in
the monolithic approximation the history of a galaxy can be described by a set
of functions of time, hierarchical histories are difficult to summarise in a
simple form, because even the identity of a galaxy is ill-defined. A galaxy is
no more a single object when viewed at different times but the ensemble of its
progenitors, all of which need to be taken into account for a correct
characterisation of the stellar population of the final object.  It is also
interesting to note that although the merger trees of these central galaxies
have a very large number of branches, only a few of these contribute
significantly to the final mass of the system (those shown as symbols which
have mass larger than $10^{10}\,{\rm M}_{\odot}\,h^{-1}$) - this relates back
to the discussion about the left panel of Fig.~\ref{fig:fig1}.

\section{Heredity}
As discussed in the introduction, a simple distinction between `heredity' and
`environment' is difficult to accommodate within the current standard paradigm
for structure formation.  Although early numerical studies found little
dependence of halo properties on environment \citep[e.g.][]{lemson99}, more
recent studies have come to different conclusions.  Taking advantage of high
resolution simulations of structure formation, recent studies have demonstrated
that halo properties like concentration, spin, shape, and internal angular
momentum show clear environmental trends \citep[e.g.][]{avila05}.  Haloes in
high density regions form statistically {\it earlier} and a higher fraction of
their mass is assembled in major mergers compared to similar mass haloes in
lower density regions \citep{gao05,maul06}.  Clearly, this is bound to leave an
`imprinting' on the galaxies that inhabit different regions today.  

\begin{figure}[!ht]
 \plotone{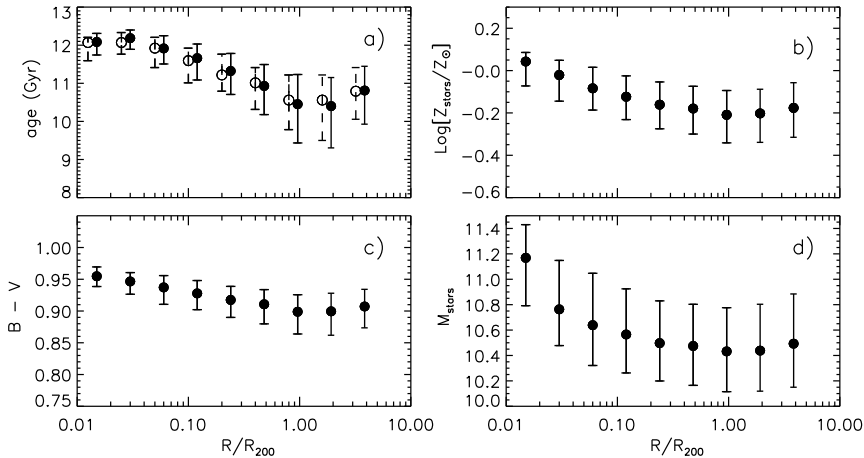}
 \caption{Median luminosity--weighted age (panel a), stellar metallicity (panel
  b), B$-$V colour (panel c), and stellar mass (panel d) for model elliptical
  galaxies in dark matter haloes with mass $> 8\times10^{14}\,{\rm M}_{\sun}$
  as a function of distance from the cluster centre.
  \citep[From][]{delucia06}.}
\label{fig:radial}
 \end{figure}

Fig.~\ref{fig:radial} shows the median properties of model elliptical galaxies
as a function of distance from the cluster centre.  Data in
Fig.~\ref{fig:radial} show that galaxies closer to the centre are on average
older, more metal rich, and redder than galaxies at the outskirts of these
clusters.  Data in panel (d) show that this trend is partly driven by mass
segregation.  A radial dependence of galaxy properties is, however, also a
natural consequence of the fact that mixing of the galaxy population is
incomplete during cluster assembly. This implies that the cluster--centric
distance of the galaxies is correlated with the time they were accreted onto a
larger structure, with galaxies closer to the centre being accreted earlier
than those residing at the outskirts \citep{diaferio01,gao04}.  If the
accretion in a larger system is associated with suppression of star formation,
the longer a galaxy is a satellite, the older its stellar population is.

The assembly history of dark matter haloes plays therefore some role in
determining the observed environmental trends.  Little work, however, has been
devoted to quantify explicitly this dependence (see Maulbetsch et al. 2006 for
a first attempt in this direction).

\section{Conclusions}
Are the observed environmental trends of galaxy properties determined very
early on (`heredity' or `nature' hypothesis) or the result of processes that
have operated during the galaxies life-times (`environment' or `nurture'
hypothesis)?  The perhaps unsurprising answer to this question is that both
necessarily play a role.  Only a few of the physical processes that I have
discussed in Sec.~2 are explicitly included in modern semi-analytic models of
galaxy formation.  Models based on merger trees extracted from numerical
simulations represent an interesting tool to be used in this context, because
they intrinsically take into account the dependencies of halo assembly history
and properties on environment.  A combination of data from the new generation
of large surveys, both in the local Universe and at high redshift, with insight
gained from modern numerical simulations of structure formation will provide an
important key to deciphering the relative importance of heredity and
environment in determining the observed environmental trends of galaxy
properties.

\acknowledgements 
I wish to thank the organisers of the conference for the invitation and for
having provided a very stimulating environment for discussion.

\end{document}